# Programmable superconducting neuron with dual-timescale plasticity for ultra-efficient neuromorphic computing

**Muen Wang, Shucheng Yang, Yuxiang Lin, Yuntian Gao, Xue Zhang, Xiaoping Gao, Minghui Niu, Huanli Liu, Yikang Wang, Wei Peng & Jie Ren***

E-mails: jieren@mail.sim.ac.cn (Jie Ren)

## ABSTRACT

The escalating energy demands of artificial intelligence pose a critical challenge to conventional computing. Combining the efficiency of event-driven, in-memory neuromorphic architectures with the ultra-high speed and low power dissipation of superconducting circuits offers a promising solution to energy-efficient computing. However, such a solution remains elusive due to the lack of a superconducting neuron that simultaneously supports programmability, local memory, and multi-timescale plasticity. Here, we introduce a programmable Josephson-junction-based leaky integrate-and-fire (LIF) neuron that features intrinsic static memory and precise programmability by encoding neuronal parameters directly in the bias current. This neuron is also capable of dual-timescale plasticity: picosecond-scale short-term modulation of spike transmission and long-term weight retention exceeding $10^4$ seconds, facilitating both rapid temporal adaptation and robust weight storage. It can operate up to 45 GHz with femtojoule-level energy dissipation per spike, and supports 10 somatic threshold levels and 20 synaptic states. Furthermore, we demonstrate a crossbar-based spiking neural network (SNN) circuit utilizing this neuron, which achieves outstanding performance across multiple tasks. System-level evaluations demonstrate a 97.48% classification accuracy on the MNIST dataset and a projected energy efficiency of 93,184 GSOPs $W^{-1}$, outperforming state-of-the-art CMOS neuromorphic systems by at least 144 times. By fusing computation, memory and plasticity into a single superconducting unit, our work paves the way for the next generation of ultrafast, energy-efficient neuromorphic computing.

## INTRODUCTION

The rapid growth of artificial intelligence (AI) continues to challenge the capabilities of conventional computing architecture built on complementary metal-oxide-semiconductor (CMOS) platforms in reducing energy consumption and latency of computation and in minimizing data movement by performing computations near memory[1]. Neuromorphic computing, which is a bio-inspired energy-efficient information processing solution, has emerged as a very strong alternative for AI infrastructure by taking advantages of event-driven operation and in-memory computation[2]. Modern CMOS neuromorphic processors, such as TrueNorth[3], Loihi[4] and Tianjic[5], have demonstrated impressive levels of parallelism and event-driven efficiency, yet they remain fundamentally limited by the memory–computation separation inherent to Von Neumann architectures. Consequently, even with event-driven architectures, emulating the rich dynamics of biological synapses and neurons still requires a massive number of CMOS transistors, resulting in significant area overhead and elevated power dissipation. Furthermore, CMOS circuits are typically confined to operating frequencies below a few GHz[6], since scaling clock speeds further results in prohibitive power dissipation and associated thermal bottlenecks that impede effective heat extraction. As a result, achieving high compute throughput and energy-efficient AI infrastructure requires a new class of hardware that unifies the following features: ultralow energy consumption, ultrafast dynamics, and learning capability.

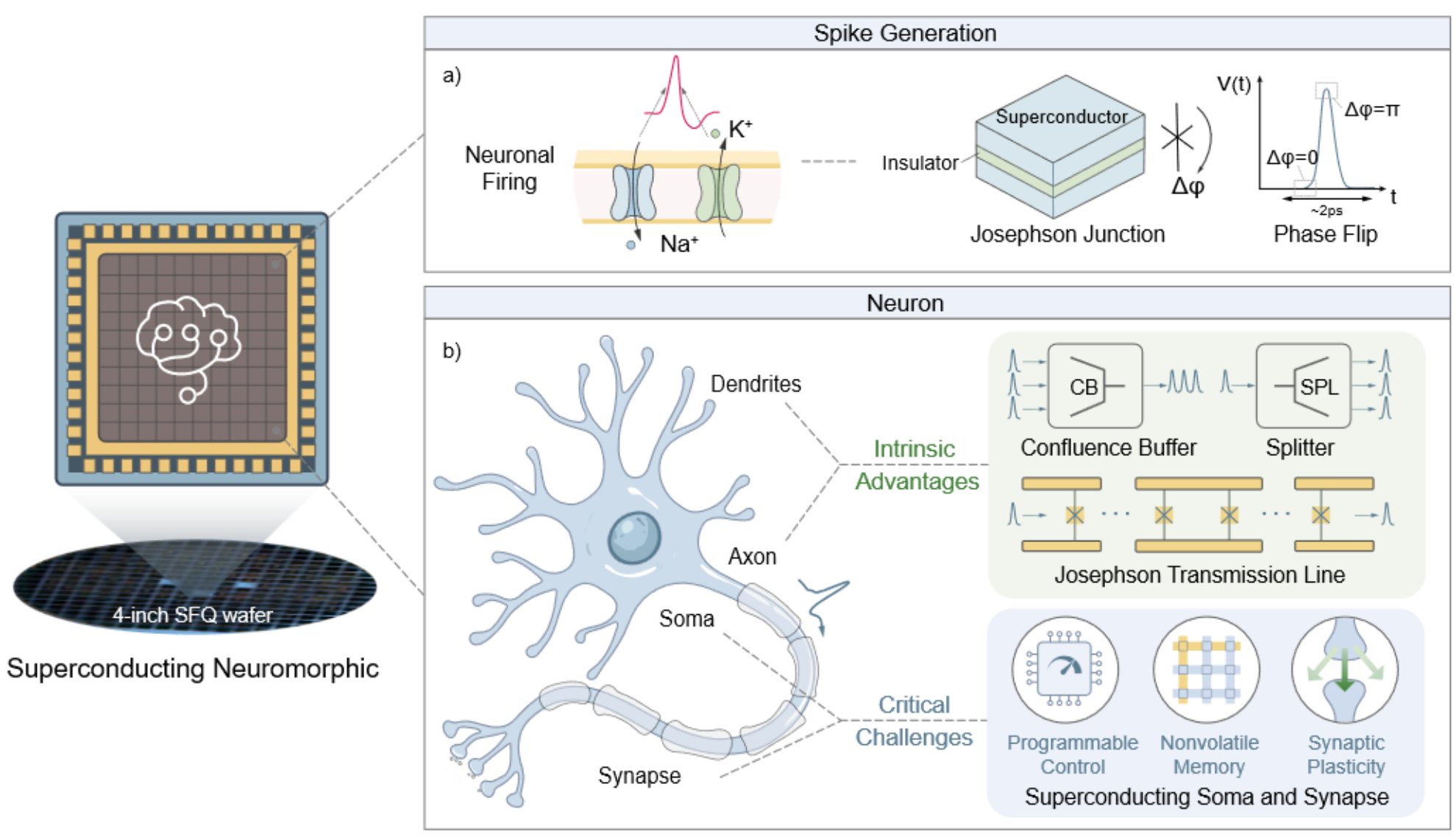

**Fig. 1 | Biomimetic mapping of neuronal dynamics and morphology to superconducting single-flux-quantum circuits. a,** Comparison of the fundamental spike generation mechanisms. The biological action potential is driven by the transmembrane influx and efflux of sodium ($Na^+$) and potassium ($K^+$) ions. In the superconducting domain, the voltage pulse V(t) is generated via a $2\pi$ phase flip. **b,** Structure and functional components of a biological neuron and their corresponding proposed superconducting circuit elements.

Superconducting circuits features ultrafast and natural pulse-based dynamics (Fig. 1a), giving rise to an intrinsic information processing similarity between superconducting circuits and biological neuron networks[7]. A biological neuron can be abstracted into four functional components: dendrite, soma, axon, and synapse (Fig. 1b). Among them, the functionalities of biological axon and dendrite can easily be emulated by low-loss Josephson transmission line (JTL) that enables long-distance pulse propagation, and by splitter (SPL) and confluence buffer (CB) that fanout and merge pulses[8], respectively. Superconducting devices such as Josephson junctions (JJ)[9–13], superconducting nanowire[14], quantum phase-slip junctions (QPSJ)[15], and magnetic Josephson junctions (MJJ)[16] can partially emulate the behaviors of soma and synapse, including membrane integration, threshold firing and nonvolatile memory, while achieving energy consumption down to the attojoule scale.

Despite these advances, experimental progress toward scalable, programmable neuromorphic systems has been limited. Most existing implementations focus on fixed-function circuits and small-scale prototypes with hardwired synaptic weights, lacking the flexibility required for diverse tasks[17–21]. This is primarily due to the difficulty in realizing the following functions within a single superconducting neuron: 1) near-memory or in-memory computing capabilities for minimizing the latency and energy overheads associated with data movement, 2) fine-grained programmability for network reconfiguration, and 3) coexisting multi-timescale plasticity to support both rapid adaptation and long-term memory. Prior studies on superconducting neurons have yet to demonstrate the above comprehensive capabilities in a single neuron.

Here we propose a practical solution, named as SPINIC (***S****uperconducting* ***P****rogrammable spl****I****king* ***N****euromorphic* ***I****ntegrated* ***C****ircuit*) by combining high-speed and low-power operation with programmability, intrinsic memory, and multi-timescale plasticity. The underlying building block of our proposed SPINIC is a programmable leaky integrate-and-fire (LIF) neuron that encodes its somatic firing threshold and synaptic weights directly into tunable bias currents of the JJs, achieving programmability without complex on-chip digital control. This mechanism also simultaneously enables in-memory computing by storing neuronal parameters in situ as persistent bias currents-a fundamental branch out from traditional superconducting logic and memory paradigms. Leveraging the unique dynamics of Josephson-junction circuits, our synapse exhibits dual-timescale plasticity: it supports picosecond-scale ($10^{-12}$ s) short-term modulation for rapid, adaptive learning and long-term weight changes stable over $10^4$ seconds for enduring memory retention. Furthermore, we fabricated a 4 × 4 SPINIC prototype core that successfully demonstrates precise inference and inherent scalability. By further scaling the architecture to a medium-scale network (e.g., a 32 × 32 core), we comprehensively evaluated SPINIC's performance. Compared with state-of-the-art CMOS neuromorphic platforms, SPINIC delivers an energy efficiency improvement of over 144 times. These results highlight its potential to drive the development of ultra-efficient neuromorphic computing platforms.

## RESULTS

**Bio-inspired superconducting neuron** As illustrated in Fig. 2a, among the four functional components of a superconducting neuron, the soma and synapse involve great computational processing complexity and thus require sophisticated circuit implementations. Mizugaki[22] previously introduced a two-Josephson-junction circuit resembling the topology shown in Fig. 2b. However, they implemented it exclusively as a fixed-weight synapse that was hardwired with a static resistor and therefore could not be dynamically adjusted. Here, we adopt this original synapse circuit topology but reinvented it to become a compact programmable LIF soma (Fig. 2b), wherein the firing threshold is continuously tunable via an external bias current. In our proposed soma circuit, the superconducting inductor ($L$) stores quantum flux and generates loop current as membrane potential, while the resistor ($R$) induces leakage to govern the decay time constant ($\tau$) of the loop current. Our simulation results in Fig. 2c show that $\tau$ can be precise tuned as $\tau \sim L/R$. A combination of picohenry-level inductance and sub-ohm resistance leads to a $\tau$ of tens of picoseconds, enabling spike computation speed exceeding 40 GHz (See Fig. S1 for circuit details). Furthermore, our circuit exhibits reliable firing triggered by a specific threshold (defined as the number of input pulses required to elicit an output spike) and reset behaviors, enabling ultra-fast neuromorphic processing. Fig. 2d shows the circuit response under threshold = 5. This reinvented soma overcomes the limitations of original design to realize highly biomimetic LIF functionality. Crucially, it unlocks the potential for intrinsic programming of neuronal parameters by allowing the loop current ($I_{loop}$) integration and firing threshold to be modulated directly via bias currents. This bias-current-based programming methodology is described in the following section.

Synapses implement weighted signal transmission by modulating the amplitude of post-synaptic potentials. An ideal superconducting synapse should scale the strength of incoming spikes to realize programmable synaptic weights. However, amplitude-based weighting is incompatible with rapid single-flux-quantum (RSFQ) circuits, as JTLs normalize pulse amplitudes to a standard SFQ. To overcome this limitation, we implement synaptic weights through pulse-count modulation: a single incoming spike triggers a programmable number of pulse replications, where the integer pulse number defines the synaptic weight. Our proposed synapse in Fig. 2e integrates the previously described LIF soma circuit (hereafter LIF) internally to generate dynamic synaptic responses. By leveraging a non-destructive read-out (NDRO) circuit driven by an inject pump input and a feedback reset signal from the LIF, the synapse performs a controlled pulse-number replication determined by the embedded LIF threshold, which corresponds to the synaptic weight $w$ (See Fig. S1 for circuit details). The output spike frequency of the synapse is governed by the LIF, ensuring compatibility with ultrafast spike processing of soma. Fig. 2f shows the simulation of a synapse with weight $w = 4$, where a train of pump pulses spaced 23 ps apart replicates each *IN* spike into four output (*OUT*) pulses. This approach bypasses the difficulty of amplitude modulation of RSFQ logic while preserving picosecond-scale spike processing speed, enabling high-speed synaptic computation within a native superconducting framework.

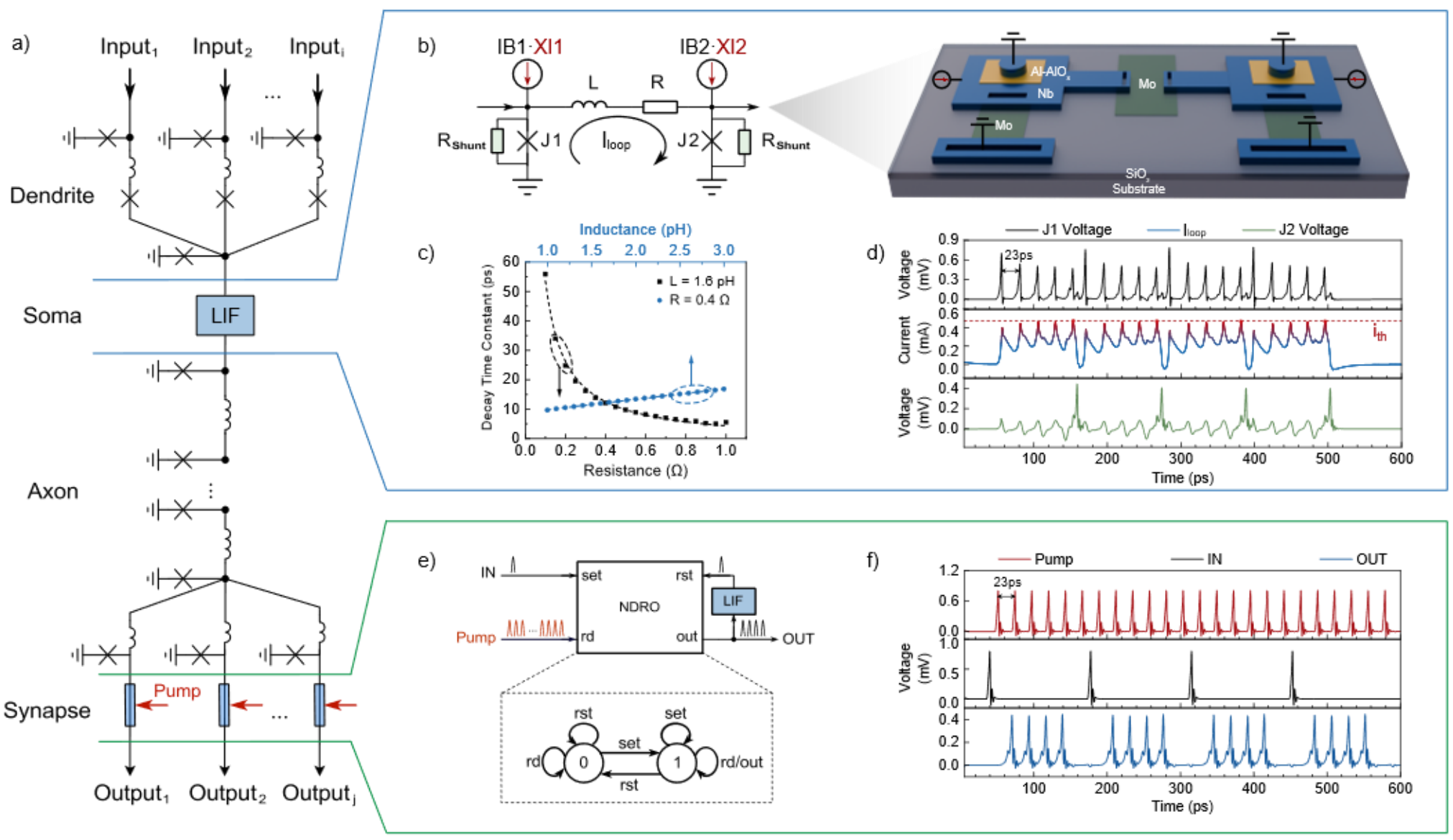

**Fig. 2 | Design of the superconducting neuron. a,** Schematic of the superconducting neuron, where the dendrite and axon are implemented using CB, JTL and SPL structures. **b,** Circuit and three-dimensional schematic of the LIF soma. Niobium (Nb) serves as the superconducting layer, Al/$AlO_x$ as the insulating layer, and molybdenum (Mo) as the resistive layer. The junction is maintained under a critically damped state for high-speed operation by shunting with a resistor to suppress. **c,** SPICE simulation of time constant vs. resistance and inductance. **d,** SPICE simulation waveforms of the LIF soma with a threshold of 5. **e,** Synapse circuit diagram and state machine of NDRO. **f,** SPICE simulation waveforms of the synapse with a weight of 4.

**Bias-current programmability** A key requirement for scalable neuromorphic hardware is the ability to store and update computational parameters directly within the processing elements themselves. Remarkably, the superconducting neuron and synapse presented here intrinsically fulfill this requirement: the DC bias currents that govern device dynamics also serve as persistent, analog memory variables. In this fashion, neuronal thresholds and synaptic weights are stored natively, without on-chip memory or refresh circuitry. By further utilizing this kind of feature, we propose a scheme to program neuronal parameters via tuning the bias currents. By introducing scaling factors *XI1* and *XI2* for *IB1* and *IB2* respectively (Fig. 2b), the LIF circuit exhibits a variable threshold. Specifically, *XI1* modulates the integration of the loop current from input pulses, while *XI2* sets the firing threshold by fine-tuning the bias point of $J_2$. As shown in Fig. 3a, our numerical simulation demonstrates that varying *XI1* and *XI2* from 0.5 to 1.5 yields ten discrete soma thresholds from 1 to 10, thereby realizing superconducting neurons with programmable states (see theory calculation in Methods). This finding is also confirmed in experiments using the soma and synapse circuits fabricated with the SIMIT-Nb03P process[23] (see Fig. S2 for experimental setup). The fabricated circuit achieves also 10 distinct threshold levels, corresponding to firing thresholds from 1 to 10. The close correspondence between numerical simulations and experimental results confirms the accuracy of the LIF model implementation and demonstrates precise, reproducible control over somatic behavior through bias-current programming.

For our synapse, a control parameter resolution of at least 4 bits is targeted, which is considered sufficient for low-precision spiking neural network (SNN) computations[24]. To fulfill our design requirement, an additional pump bias current channel (*IB3*, with a scaling factor of *XI3*, as highlighted in red in Fig. S1b) is added adjacent to the LIF unit. This design is validated in experiments with results demonstrating precise control of the LIF's trigger current, enabling synaptic weight outputs ranging from 1 to 20, as shown in Fig. 3b. Our results further suggest that increasing any of the three bias currents reduces the threshold for flux integration, thereby increasing the probability of trigger events and effectively decreasing the synaptic weight. The mechanism behind this continuous and analog tuning behavior arises from the fundamental phase dynamics of JJs: altering bias current reshapes the potential landscape of JJs, enabling precise and reversible control over the likelihood of SFQ pulse generation. This mechanism enables precise in situ programming of synaptic weight with minimal hardware overhead. By integrating storage and computation parameters into the same circuit, the system avoids data movement overheads and realizes an intrinsic form of computing-in-memory (CIM).

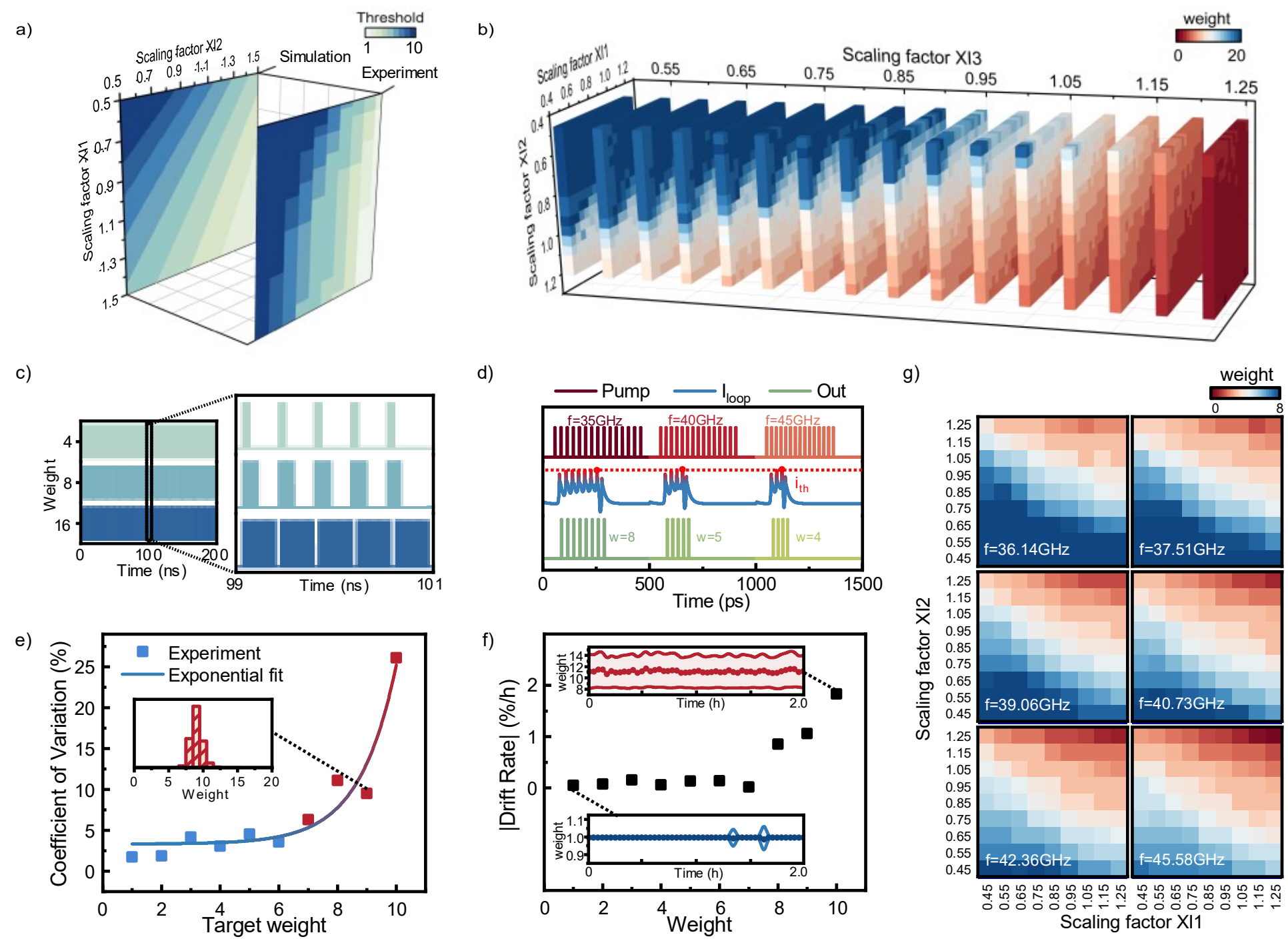

**Fig. 3 | Design and experimental characterizations of the superconducting neuron with programmability and dual-plasticity. a,** Numerical simulation and experimental results showing the impact of the bias current scaling factors on the soma threshold. **b,** Experimental result showing the relationship between the bias current scaling factors and the synaptic weight. **c,** Simulation of long-term plasticity. The results with weights of 4, 8, and 16 are shown respectively, with corresponding bias-current scaling factors (*XI1, XI2, XI3*) being: (1.2, 1.2, 1.0), (1, 1, 0.8), and (0.95, 0.85, 0.55). **d,** Simulation of short-term plasticity. **e,** Coefficient of variation (*CV*) of synaptic weights. The inset displays the distribution of experimental results for a target weight of $w$=9. **f,** Drift rates (*DR*) of synaptic weights. The insets illustrate the temporal evolution of weights for the most ($w$ = 1) and least ($w$ = 10) stable cases. **g,** STP characteristics of the synapse. For clarity, *XI3* is fixed at 1.0 here.

**Dual-timescale plasticity of synapse** The intrinsic plasticity of synapses enables artificial neurons to dynamically adjust their excitability, mimicking the adaptive abilities of biological neurons. This functionality optimizes information processing efficiency and significantly enhances learning capacity[25]. Unlike conventional memristive synapses that rely on conductance modulation[26–28], our synapse achieves plasticity by modulating both the integration and the firing current within the LIF feedback loop. Long-term plasticity (LTP) is realized via varying the bias current of JJs in the LIF feedback loop, allowing stable weight adjustment and thus supporting learning mechanisms. Fig. 3c presents simulation results for three distinct weight states over a 200 ns duration with a zoom-in view for detailed waveform. Although the simulation timespan is constrained by the high computational cost of analog transient analysis, the observed long-term stability effectively validates the circuit's dynamic robustness. Short-term plasticity (STP) arises from the temporal dynamics of loop current accumulation, where the frequency of an input pulse train directly governs the extent of current decay between spikes. As illustrated in Fig. 3d, increasing the input pulse frequency from 35 GHz to 40 GHz and then to 45 GHz reduces the weight from 8 to 5 and finally to 4. The simulation result shows that altering the input pulse frequency, can instantly increase or decrease the decay of the loop current during the temporal interval between two consecutive input pulses, thereby adjusting the number of output pulses. Simulations with input frequencies ranging from 35 to 45 GHz (equivalent to pulse periods of 28.6–22.2 ps) reveal substantial changes in weight. This indicates that modulating the pulse period by mere picoseconds is sufficient to effectively tune the short-term excitability of the synapse.

To investigate LTP of our synapse experimentally, we recorded weight states encoded by bias currents over a continuous 20-hour period. Ten distinct weight states were each evaluated over 10,000 high-frequency hardware test cycles. The extracted coefficient of variation (*CV*) and drift rate (*DR*), shown in Fig. 3e and Fig. 3f, are then used to quantify the stability. The results indicate that *CV* increases with the target weight state, exhibiting an approximately exponential trend. This behavior arises from the fact that the accumulation of input pulses diminishes the margins between adjacent loop-current peaks (Fig. S3), resulting in an exponential reduction in current resolution. Based on a standard 5% *CV* tolerance for typical neuromorphic systems[29,30], we achieve six distinguishable stable states (i.e., $w \leqslant 6$). Notably, all ten weights exhibit negligible temporal drift ($|DR| < 2\%$), confirming the stability of the programming currents and the robustness of our implemented circuits[31]. STP was characterized by measuring synaptic weight responses to different input pulse frequencies. Fig. 3g shows the relationship between input frequency (36.14-45.58 GHz) and synaptic weight at six representative frequency points (See high-frequency calibration in Methods). Increasing the input frequency reduces the synaptic weight, consistent with our previous analysis. The average weight reduction is approximately 0.22 per GHz, demonstrating frequency-modulated STP behavior. By integrating rapid modulation and long-term weight storage within a single superconducting circuit, our architecture realizes dual-timescale plasticity with minimal hardware overhead. This capability enables scalable bio-inspired learning on an ultra-efficient superconducting platform, combining ultrafast responses beyond CMOS and memristive devices with robust long-term retention.

**Network architecture design and a small-scale experimental validation** To evaluate system-level behavior of the proposed superconducting neuron in SNNs, we implemented a superconducting neuromorphic architecture called SPINIC, as mentioned above. As shown in Fig. 4b, the architecture adopts a crossbar structure in which input signals (*DataIn*) are distributed along each row via in-synapse SPLs and sequentially delivered to all synapses in that row. The synaptic outputs are then summed along the column direction through CBs and routed to the soma located at the bottom of each column, where membrane potential integration and spike generation occur. As illustrated in Fig. 4c, this interface design enables modular tiling of SPINIC cores, allowing the system to scale in both width and depth with minimal interconnect overhead.

Specifically, we fabricated and characterized a SPINIC prototype core with the array scale of 4 × 4 to demonstrate its scalable processing capabilities. As shown in Fig. 5a, the network circuitry (1,050 JJs) occupies approximately one-third of the total circuit area, while the remaining area is dedicated to auxiliary high-speed test structures (2,474 JJs). All parameters of SPINIC, including somatic thresholds and synaptic weights, are set through static DC bias currents applied to the corresponding JJs. This eliminates the need for any on-chip digital memory or weight-loading circuitry. Once programmed, all network states remain static under constant bias, with no refresh overhead.

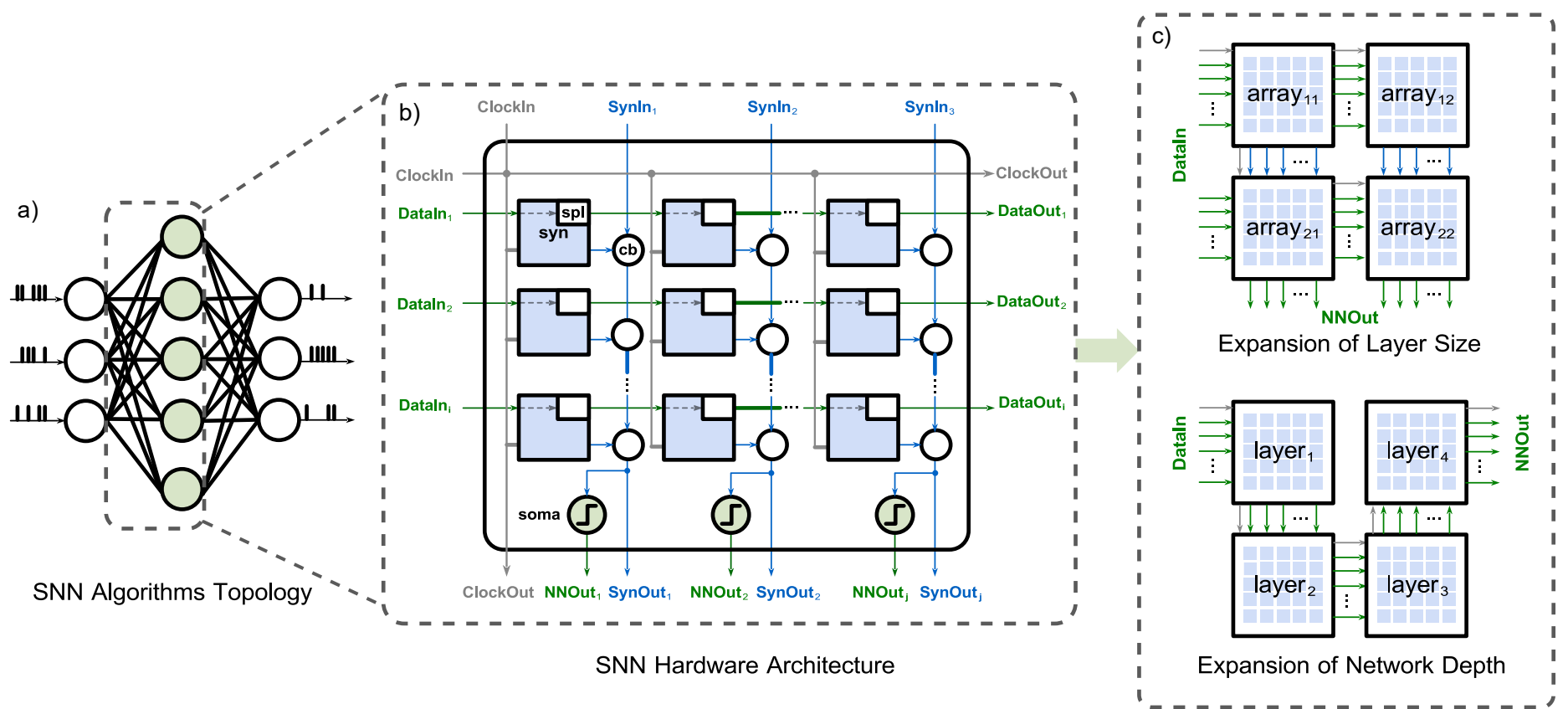

**Fig. 4 | Architecture of the SPINIC. a,** Topology of SNN algorithm. **b,** Network architecture of SPINIC. Blue boxes represent synapses and green circles denote somas. **c,** Illustration of SPINIC's expansion interconnection scheme. Two configurations are shown, one for the expansion of the layer size and another for expansion of the network depth.

In our experiments, we constructed several representative neural networks to verify the programmability of the circuits. Test datasets, including 4-pixel images, 4-pixel noisy images, and the Iris dataset (Fig. 5a), were first used to train each network offline, thereby determining the corresponding synaptic weight matrices and somatic thresholds, which were subsequently mapped to the hardware by adjusting the corresponding bias currents. To ensure accurate parameter mapping, we established a calibration relationship between the bias currents and the neuronal parameters prior to functional network testing. This was accomplished by sweeping the current scaling factors to directly measure the resulting thresholds and weights, producing a deterministic current-to-parameter lookup table for accurate programming later on. The reconfiguration time of the entire 4 × 4 network is only on the order of microseconds, which in our case is dictated by the limited update rate of our external DC bias sources[32] (a limit that in principle could be further reduced with on circuitry bias sources), demonstrating true general-purpose programmability. Fig. 5b displays representative measured waveforms from a 4-pixel image inference test, showing input spikes (Datain1–4), readout signals (Read), and output pulses (NNout1–4). Complete results across all three test sets are summarized in Fig. 5c, demonstrating full consistency with both algorithmic expectations and circuit simulations. Overall, the 4×4 SPINIC core successfully performed accurate classification on these representative datasets, confirming the feasibility and programmability of our proposed superconducting neuromorphic architecture and laying a foundation for scaling toward larger systems.

Crucially, parameter variations induced by fabrication inhomogeneity can lead to unintended weight deviations, thereby degrading the computing accuracy of the network. Therefore, verifying the consistency across spatial scales is a prerequisite for ensuring the robustness of large-scale superconducting neuromorphic systems. Therefore, to evaluate the spatial consistency of the proposed superconducting neurons, we experimentally examined the distribution of a key parameter, i.e., weight values, of the 16 synapses within the network. As shown in Fig. 5d, the on-chip distance between synaptic circuits ranges from 300 μm to 1338 μm. The weight values were extracted through bias current scanning, and a two-sample Kolmogorov–Smirnov (K-S) test was performed for statistical analysis[33] (see details in Supplementary). The synapse located at position (1,1) was used as a reference to evaluate the other 15 synapses. All other synapses exhibited very small K-S distances ($D$ < 0.15) and nearly overlapping empirical cumulative distribution functions (ECDFs) with the reference, indicating high consistency in weight distribution across the chip. This suggests excellent spatial robustness of the synaptic circuit implementation.

**System-level performance and scalability** To evaluate the computational prospect of the SPINIC architecture under hardware-intrinsic constraints, we benchmarked its inference capabilities using standard neuromorphic datasets (MNIST[34] and Fashion-MNIST[35]). A four-layer SNN model was designed, consisting of an input flattening (IF) layer and three fully connected (FC) layers with dimensions of IF(28×28)-FC(512)-FC(64)-FC(10). Given that superconducting logic inherently operates with limited parameter precision (for instance, 20 discrete weight states in our system), we rigorously evaluated the impact of the quantization effect on the network performance. As illustrated in Fig. 5e, this architecture exhibits remarkable robustness to reduced-precision constraints. Compared to full-precision baselines, the quantized SPINIC model incurs a marginal accuracy loss of only 0.21% for MNIST and 1.04% for Fashion-MNIST. This validates that the bias-current programming method in our proposed superconducting neuromorphic scheme effectively supports accurate inference despite the limited parameter bit-width of neurons.

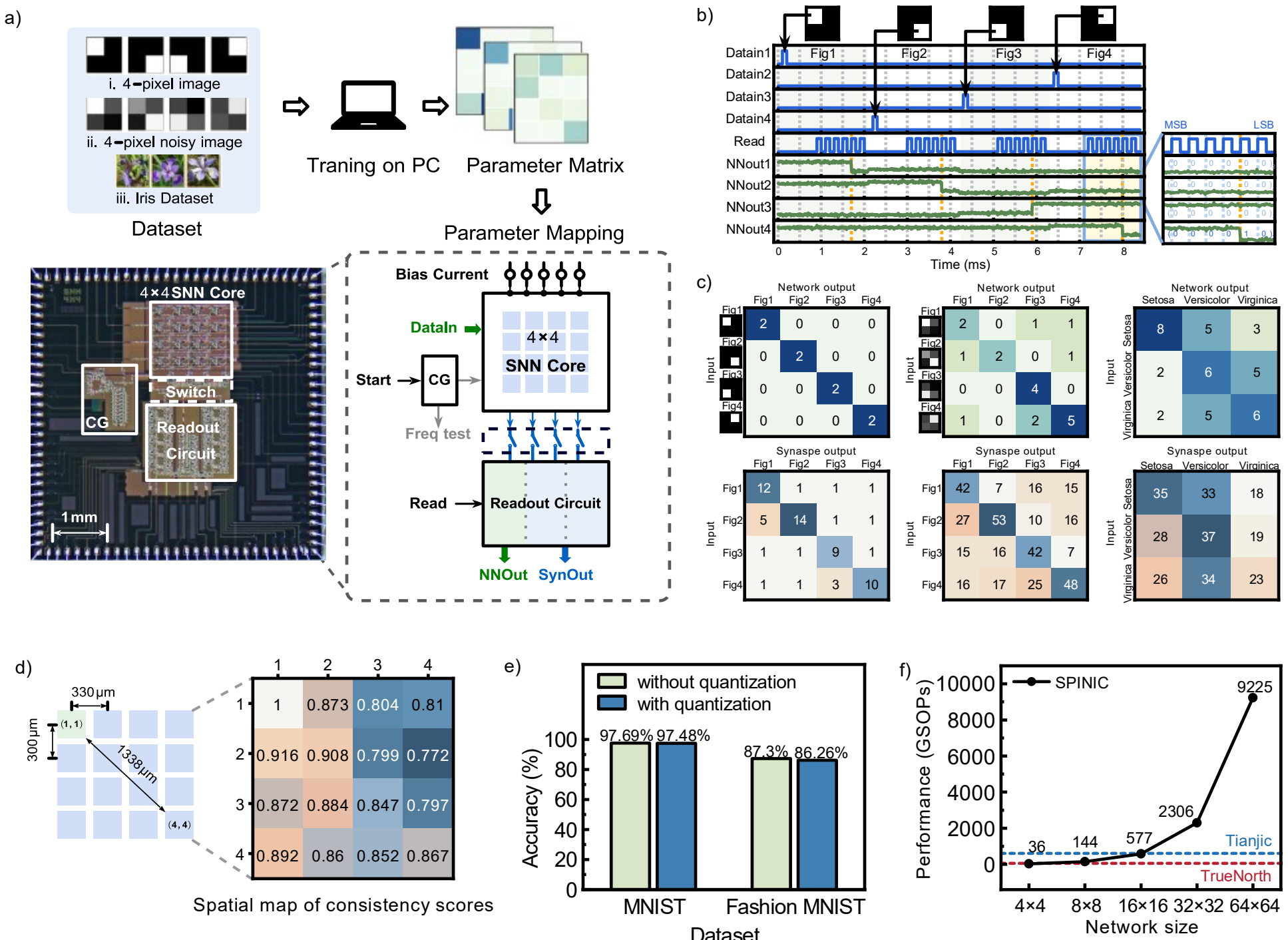

**Fig. 5 | Experimental results of the 4 × 4 SPINIC core. a,** Testing workflow, chip micrograph, and circuit schematic of the 4×4 SPINIC. **b,** Example of measured waveforms from the 4×4 SPINIC. **c,** Inference results for the three datasets. **d,** Spatial consistency validation of synaptic parameters in SPINIC. **e,** Algorithmic accuracy of SPINIC on MNIST and Fashion-MNIST. **f,** Throughput of SPINIC at different network scales and comparison against CMOS-based benchmarks.

The fundamental advantage of the SPINIC paradigm lies in its energetic regime. In superconducting SFQ circuits, total power consumption consists of static and dynamic components[36]. Based on the measured device characteristics of SPINIC, the energy consumption for a single synaptic operation here is only 3.21 fJ/SOP. This femtojoule-level operation represents a reduction of approximately three orders of magnitude compared to state-of-the-art CMOS neuromorphic platforms (Table 1). We further project the system-level scalability of SPINIC by extending our experimentally characterized 4 × 4 core to larger integration scales (refer to Supplementary for detailed calculation). While current superconducting fabrication processes have demonstrated circuits with up to $1.03 \times 10^6$ JJs[37], practical integration is often constrained by flux trapping and bias distribution overheads. Under a conservative estimation of an integration scale of less than $10^6$ JJs, a medium-scale 32×32 SPINIC core (requiring ~ 67,500 JJs) should be feasible with a total power consumption of roughly only 24.75 mW. In order to quantify its compute throughput, the metric of synaptic operations per second (SOPS), first introduced by IBM[38], is used. As illustrated in Fig. 5f, we evaluated the peak performance of SPINIC cores ranging from 4 × 4 to 64 × 64 in size and compared that with advanced CMOS-based designs. Noticeably, the 32 × 32 SPINIC core is capable of a peak throughput of 2,306 GSOPS, surpassing that of large-scale CMOS neuromorphic processors. Due to the ultralow switching energy and minimal delay of SFQ circuits, our SPINIC core could achieve 93,184 GSOPS $W^{-1}$, an energy efficiency performance level at least 144 times better than the CMOS counterparts. Compared with the recently proposed superconducting SUSHI chip with similar JJ counts, SPINIC is 10 times better in energy efficiency, attributed to its simplified and highly optimized neuron architecture. In practice, a fair comparison should take into account the cryogenic cooling power penalty of our system. Even considering an exaggerated cooling power factor of 300 times, our system is still capable of 311 GSOPS $W^{-1}$. With further optimization by adopting energy-efficient RSFQ (ERSFQ) technology, the static power dissipation that dominates in current designs can be virtually eliminating. The resulted system efficiency is projected to reach 8,962 GSOPS $W^{-1}$, a 19-fold advantage over advanced CMOS implementations.

**Table 1 | Performance comparison between the SPINIC and previously reported neuromorphic hardware**

| | Year | Implementation | Synaptic width | GSOPS | J $SOP^{-1}$ | GSOPs $W^{-1}$ | |
|---|---|---|---|---|---|---|---|
| | | | | | | Wo.C.[a] | W.C.[b] |
| SpiNNaker[39] | 2013 | Soft-core-based | - | 0.064 | 11.3 n | 0.064 | 0.047 |
| TrueNorth[38] | 2015 | ASIC | 1 bit | 58 | 26 p | 400 | 292 |
| Loihi[4] | 2018 | ASIC | 1 to 9 bit | - | 23.6 p | 42 | 31 |
| Tianjic[5] | 2020 | ASIC | 8 bit | 608 | 1.54 p | 649 | 474 |
| Spiker[40] | 2022 | FPGA-based | 16 bit | - | 41 n | 0.024 | 0.018 |
| Darwin3[41] | 2024 | ASIC | 1/2/4/8/16 bit | - | 5.47 p | 183 | 134 |
| SUSHI[19] | 2023 | Superconducting Devices | 1 bit | 1355 | - | 32,336 [c] | 108 [c] |
| SPINIC | 2026 | Superconducting Devices | 4 to 5 bit | 2,306 | 3.21 f | 93,184<br>314,000 [c]<br>2,693,949 [d] | 311<br>1046 [c]<br>8,962 [d] |

[a] Without cooling cost; [b] Cooling overhead is estimated at 1.37× for CMOS chips (global data center average in 2020)[10] and intentionally exaggerated to 300× for superconducting chips to highlight performance under highly unfavorable cooling assumptions; [c] The estimation does not include the layout and interconnect overhead required for practical superconducting implementation; [d] Based on the ERSFQ technology.

## CONCLUSIONS

We have demonstrated a fully programmable, scalable superconducting neuromorphic architecture that unifies the extreme speed of RSFQ logic with the adaptability required for intelligent processing. By identifying bias currents as a new degree of freedom for parameter encoding, SPINIC overcomes the long-standing challenge of implementing local memory in standard superconducting circuits. This innovation enables the construction of neurons that are not only orders of magnitude faster and more efficient but also capable of complex, multi-timescale learning. System-level demonstrations show that a SPINIC core operates at frequencies up to 45 GHz with an energy consumption of 3.21 fJ per synaptic operation and achieves a projected energy efficiency of 93,184 GSOPS $W^{-1}$, outperforming state-of-the-art CMOS neuromorphic processors by more than two orders of magnitude. These results establish SPINIC as a promising platform for ultrafast and energy-efficient neuromorphic computing and provide a viable pathway toward scalable superconducting intelligence hardware.

## METHODS

**Theory calculation of bias-current programming** The LIF output is decided under the condition whether the current through $J_2$ (Fig. 2b) exceeds its critical value. According to Kirchhoff's current law, the triggering condition can be expressed as,

$$I_{B2}+I_{loop}>I_{C2}, \tag{1}$$

where $I_{loop}$ exhibits step increment with input pulses and decays over time. Since $J_2$ triggers at the peaks of $I_{loop}$, we simplify the condition to consider only the peak loop current $I_{loop}^{pk}$,

$$I_{B2}+I_{loop}^{pk}(N)>I_{C2} \tag{2}$$

$$I_{B2}+I_{loop}^{pk}(N\text{-}1)<I_{C2}. \tag{3}$$

The peak loop current after $N$ pulses, $I_{loop}^{pk}(N)$, can be computed recursively as,

$$I_{loop}^{pk}(N)=I_{loop}^{pk}(N\text{-}1)\text{-}I_{dec}+I_{inc}, \tag{4}$$

where $I_{inc}$ is the current increment per pulse and $I_{dec}$ is the current decrement per pulse. Roughly to the first order estimation, both $I_{inc}$ and $I_{dec}$ decrease simultaneously slightly by the same amount as more flux accumulates. Specifically, the reduction in $I_{inc}$ stems from incomplete flux transfer in $J_1$, while the decrease in $I_{dec}$ arises from changes in the loop's effective inductance that alter the decay time constant $\tau$. Since the corresponding impact of simultaneous decrement of $I_{inc}$ and $I_{dec}$ on the above peak loop current after $N$ pulses, $I_{loop}^{pk}(N)$, nearly cancel, we approximate $I_{inc}$ and $I_{dec}$ as constants. The peak current after $N$ pulses then is simplified to the following,

$$I_{loop}^{pk}(N)=NI_{inc}\text{-}(N\text{-}1)I_{dec}. \tag{5}$$

The flux released by $J_1$ induces $I_{inc}$ on $L$, which is also partially modulated by $IB_1$. During the pulse interval $\frac{1}{f}$, the current decays by $I_{dec}$,

$$I_{inc}=a+bI_{B1} \tag{6}$$

$$I_{dec}=I_{inc}(1\text{-}e^{-\frac{1}{\tau f}}) \tag{7}$$

Substituting into the firing condition yields

$$\left(\frac{I_{C2}\text{-}I_{B2}}{a+bI_{B1}}\text{-}1\right)e^{\frac{1}{\tau f}}+1<N<\left(\frac{I_{C2}\text{-}I_{B2}}{a+bI_{B1}}\text{-}1\right)e^{\frac{1}{\tau f}}+2 \tag{8}$$

This expression suggests that the LIF threshold can be tuned by multiple circuit parameters. In practice, we focus on controllable post-fabrication parameters, $I_{B1}$, $I_{B2}$, and the input pulse frequency $f$. Based on this principle, we propose a scheme to program neuron parameters via bias currents by introducing scaling factors *XI1* and *XI2* for $I_{B1}$ and $I_{B2}$.

**Fabraication process of SPINIC** The SPINIC chip and the neuron circuits were fabricated using the SIMIT-Nb03P

superconducting integration process. This process employs 4-inch surface-oxidized silicon wafers as the substrate, with high-quality $SiO_2$, deposited by plasma-enhanced chemical vapor deposition (PECVD), serving as the insulation layer, Ti/Au acting as the contact layer, and Nb superconducting layer prepared by magnetron sputtering. Highly stable and reproducible Josephson junctions were achieved by depositing Nb/Al–$AlO_x$/Nb trilayer films without breaking the vacuum, followed by junction definition via deep ultraviolet (DUV) stepper lithography and inductively coupled plasma (ICP) etching.

**Experimental setups for superconducting circuits** The processed 4-inch wafer was first diced into 5×5 $mm^2$ chips, which were then wire-bonded to a printed circuit board (PCB) using highly conductive aluminum wires to minimize electrical losses. The PCB was mounted into a 116-channel cryogenic probe, with electrical connections established between the board and the probe electrodes. The probe was immersed in a liquid helium dewar, cooling the chip to an operating temperature of 4.2 K. Electrical characterization was performed via three DB50 cables connecting the probe to an OCTOPUX-4FH/128H multi-channel automated test system—a versatile low-frequency measurement platform comprising 128 high-precision DAC and ADC channels, along with a digital voltmeter (DVM) offering 1 μV resolution. System control and circuit testing were automated using a custom MATLAB script to coordinate all OCTOPUX channels.

**On-chip high-frequency testing method** The on-chip clock generator (CG) frequency was adjusted via changing its bias current. Higher bias currents accelerate junction switching and reduce pulse transmission delay, thus increasing the CG frequency. The actual CG frequency was determined from the average voltage across the JJs using Faraday's law. When SFQ pulses pass a junction at frequency $f$, the junction voltage satisfies $V=f\Phi_0$, where $\Phi_0$is the superconducting flux quantum. Therefore, by measuring the average voltage $\overline{V_{CG}}$, the CG frequency can be determine using the following formula,

$$f=\frac{\overline{V_{CG}}}{\Phi_0}$$

Notice, the average voltage across the junction is measured using the above mentioned high-precision DVM of the OCTOPUX system.

**SPICE Simulation Setup** All circuit-level simulations were performed using in-house developed JSICsim[42], a SPICE-compatible simulator tailored for superconducting integrated circuits. The structural netlist was generated from the schematic of the neuron unit and its testbench, preserving the hierarchical connectivity of the physical design. To capture the dynamic switching behavior of JJs, we specified timing and triggering conditions using SFQ Hardware Description Language (SFQ HDL). This behavioral description was co-exported with the netlist and translated into Verilog-A models compatible with JSICsim as needed, enabling mixed structural-behavioral transient simulations.

Functional correctness was first verified under nominal process conditions. Subsequently, we quantified circuit robustness through operating margin extraction. Specifically, we employed a bisection-based search algorithm to determine the tolerance limits of three key process parameters: the bias current scaling factor (*XI*), Josephson critical current density scaling factor (*XJ*), and inductance scaling factor (*XL*). For each parameter, the algorithm iteratively narrowed the interval between a known functional point and a failure point until the boundary of correct SFQ pulse propagation was identified within a predefined resolution. This approach efficiently maps the process window, in which the circuit remains operational despite fabrication-induced variations.

## MISCELLANEA

**Supplementary material** Supplementary material associated with this article can be found in the supplementary file.

**Acknowledgments** This work was supported by the National Natural Science Foundation of China under Grant No. 62571520, the Autonomous deployment project of State Key Laboratory of Materials for Integrated Circuits No. SKLJC-Z2024-A03.

**Declaration of Competing Interest** The authors declare no competing interests.

## Supplementary Materials

### Working Principle of Soma and Synapse Circuit

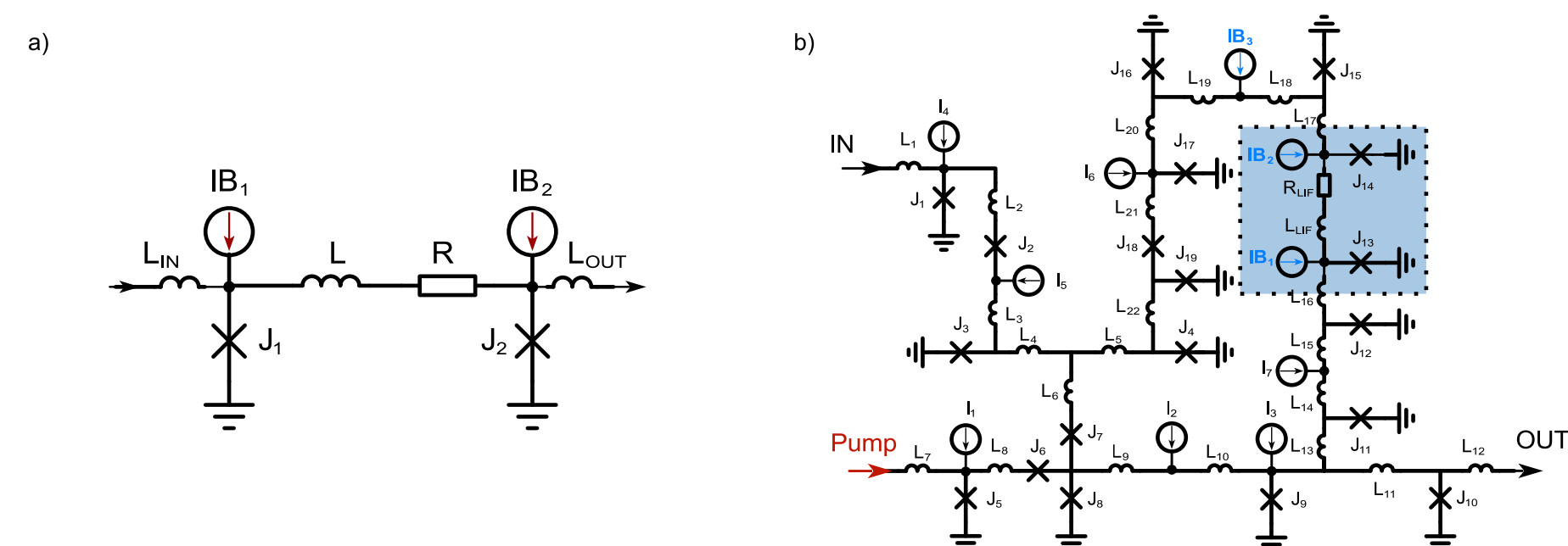

**Supplementary Fig 1 | Schematic of the superconducting neuron a** LIF soma circuit schematic. The values of the parameters are： $J_1$ = 275 μA, $J_2$ = 370 μA, $L_{IN}$ = 0.62 pH, $L$ = 2.76 pH, $L_{OUT}$ = 1.44 pH, $IB_1$ = 187.5 μA, $IB_2$ = 100 μA, $R$ = 0.3 Ω. **b** Synapse circuit schematic. The blue region highlights the LIF used for counting output pulse and resetting signal generation. Three sets of current sources, marked in blue, represent dedicated bias channels for synaptic weight modulation. The values of the parameters are： $J_1 = J_5 = J_{10} = J_{11} = J_{12} = J_{15} = J_{16} = J_{17}$ = 250 μA, $J_2$ = 290 μA, $J_3 = J_{18}$ = 306.25 μA, $J_4$ = 210 μA, $J_6 = J_8$ = 393.75 μA, $J_7$ = 165 μA, $J_9$ = 325 μA, $J_{13}$ = 275 μA, $J_{14}$ = 370 μA, $J_{19}$ = 237.5 μA, $L_1 = L_{12} = L_{14} = L_{15} = L_{18} = L_{19} = L_{21} = L_{23}$ = 2.0 pH, $L_2$ = 1.59 pH, $L_3$ = 0.92 pH, $L_4$ = 3.25 pH, $L_5$ = 1.68 pH, $L_6$ = 0.64 pH, $L_7$ = 2.7 pH, $L_8$ = 2.29 pH, $L_9 = L_{10}$ = 1.2 pH, $L_{11}$ = 3.3 pH, $L_{13}$ = 3.1 pH, $L_{16}$ = 1.88 pH, $L_{LIF}$ = 2.76 pH, $L_{17}$ = 3.15 pH, $L_{20}$ = 4.22 pH, $L_{22}$ = 1.07 pH, $I_1$ = 195 μA, $I_2$ = 280 μA, $I_3$ = 132.5 μA, $I_4$ = 105 μA, $I_5$ = 170 μA, $I_6$ = 215 μA, $I_7 = IB_3$ = 350 μA, $IB_1$ = 140 μA, $IB_2$ = 80 μA, $R_{LIF}$ = 0.3 Ω.

We proposed a compact soma circuit comprising two Josephson junctions ($J_1$, $J_2$), an inductor $L$, and a resistor $R$ as shown in Fig. 1a. $J_1$ and $J_2$ serve as the integration and triggering junctions, respectively. $J_1$ has a smaller critical current and is biased with $I_{B1}$, enabling rapid response to input pulses which induces a phase slip in $J_1$ and injects flux into the loop that increase the circulating current. $J_2$ has a larger critical current and a smaller bias $I_{B2}$, acting as the output junction. It fires only when the sum of the loop current and $I_{B2}$ exceeds its critical current $I_{C2}$. This mechanism reproduces the membrane potential accumulation and threshold-triggered firing observed in biological neurons. Here the inductor $L$ in the circuit loop stores magnetic flux with negligible loss, representing the membrane potential, and the inserted resistor $R$ induces temporal flux decay, defining a key LIF parameter, i.e., the leak time constant.

The superconducting synapse in Fig. 1b integrates the LIF soma circuit (hereafter LIF) in Fig. 1a internally to generate dynamic synaptic responses. Here, *IN* is the input, while *TI* serves as a "pump" signal that drives pulse multiplication. A feedback LIF loop ensures self-terminating multiplication by accumulating output pulses until the threshold is reached. Upon receiving an SFQ pulse at *IN*, flux propagates through $J_1$ into the $J_3 – J_4$ loop and is then stored, generating a clockwise circulating current that biases $J_7$ and $J_8$ to a precritical state. In this case, each pump pulse from *TI* then triggers phase slips in $J_7$ and $J_8$ sequentially. The switching of $J_7$ cancels the flux perturbation in the $J_3 – J_4$ loop, allowing the circulating current to persist. Switching of $J_8$ generates an output SFQ pulse through $J_9 – J_{10}$ and simultaneously feeds it back to the $J_3 – J_4$ loop via $J_{11} – J_{19}$. The inserted LIF circuit ($J_{13} – L – R – J_{14}$) integrates these feedback pulses and triggers once its threshold is reached. When enough output pulses are accumulated to surpass this threshold, $J_{14}$ produces a reset pulse that injects a counter-clockwise current into the $J_3 – J_4$ loop, erasing the stored flux. Subsequent *TI* pulses are then released from $J_6$ instead of propagating through $J_8$, thereby halting the multiplication process.

## High-Frequency Testing of Superconducting Neurons and Neural Networks

To characterize the high-frequency performance of the neurons, we implemented an on-chip 44 GHz clock generator (CG) using SFQ circuits. The pulses generated by the CG propagate through the soma and synapse circuits, with output signals captured by dedicated readout structures. For the soma, output pulses are counted by a superconducting binary counter (Fig. 2a and 2b), while the synaptic circuit employs an on-chip shift register (SR) to buffer output data (Fig. 2c and 2d). Both readout circuits interface with external instruments via 10 kHz low-frequency control signals, enabling reliable data extraction while avoiding bandwidth bottlenecks and excessive thermal load on the cryogenic platform.

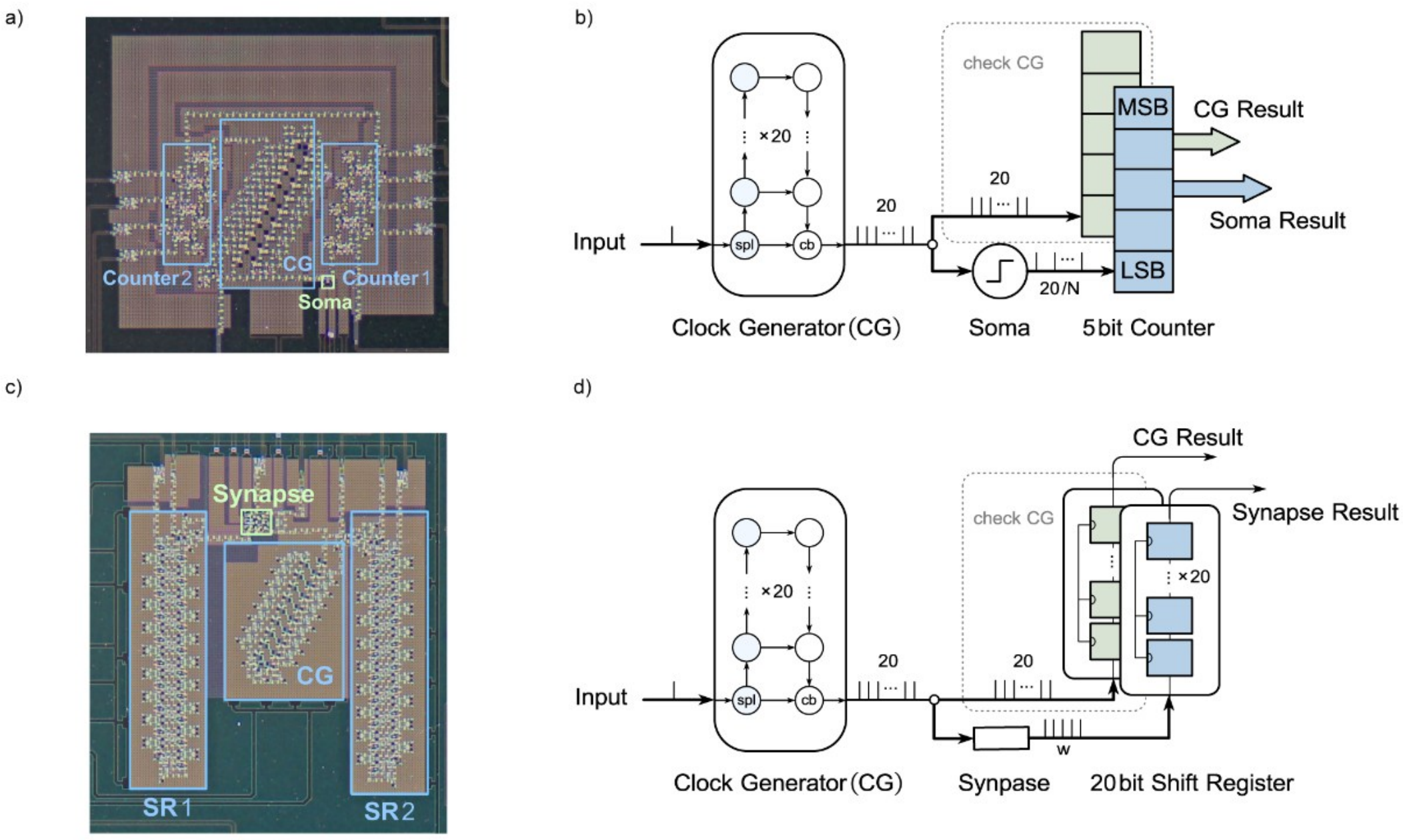

**Supplementary Fig 2 | High-frequency testing of superconducting neurons a** Micrograph of soma testing circuit. **b** Test circuit of the LIF soma. The CG produces 20 high-frequency input pulses to the soma, which generates output pulses according to its threshold value $N$, resulting in $20/N$ output spikes. The green and blue 5-bit counters record the pulse counts of the CG and the soma outputs, respectively. The green counter is used to verify whether the CG correctly produces 20 pulses and thus ensures accurate threshold statistics. **c** Micrograph of synapse testing circuit. **d** Test circuit of synapse circuit. The CG generates 20 input pulses, and the synapse produces $w$ output pulses based on its synaptic weight. The green and blue SR circuit buffer and read out the high-frequency pulses from the CG and the synapse, respectively. The green SR verifies correct pulse generation by the CG to prevent miscounting of synaptic weights.

The characterization of the SPINIC neural network follows a similar approach as the neuron-level tests. The on-chip CG supplies high-frequency signals, and network outputs are acquired via pulse counting and low-frequency readout circuits. One dedicated counter records the network outputs, while two additional counters monitor activity in the synaptic layer (pre-activation layer). To minimize on-chip resource usage, the outputs of the four output-layer neurons are counted in a time-division multiplexing manner. A switching circuit sequentially selects each column for counting while keeping others disabled, allowing complete acquisition of network outputs by cycling through all four columns.

## Evaluation of Long-Term Plasticity (LTP)

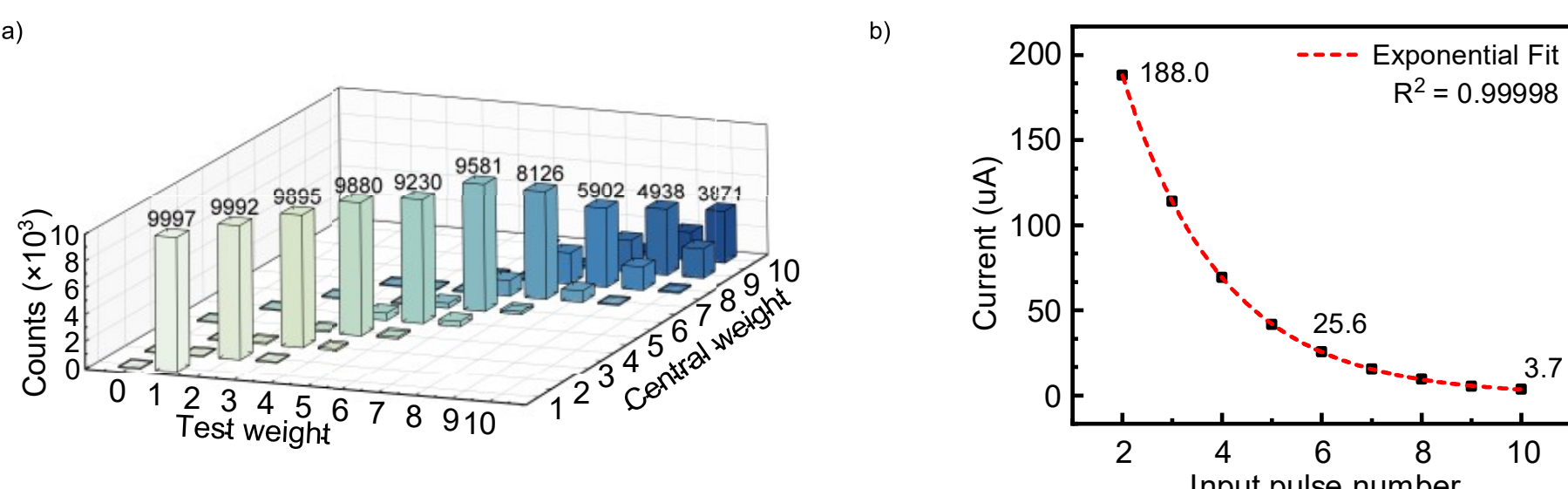

**Supplementary Fig 3 Analysis of the long-term plasticity in synapse circuits a** Statistical distribution of LTP weight parameters. The histogram represents weight distributions across 10,000 test cycles for ten target levels (1–10) **b** Peak loop current increments as a function of spike number in the LIF of synapse circuits simulated using SPICE. Black dots represent simulation results; red dashed lines denote exponential fitting results.

For LTP, synapses were continuously tested for over 20 hours to assess the stability of weight changes induced by adjustments to the bias currents. Ten weight points were tested, each with 10,000 high-frequency cycles. As shown in Fig. 3a, the statistical distribution of test results indicates that weight stability decreases as the target weight increases, following an approximately sigmoidal trend. For weights ≤ 6, their coefficient of variation (CV) remains below 5%. In addition, lower weights exhibit high stability, whereas higher weights produce approximately normal distributions centered around the target weights, with increasing standard deviation. This behavior arises from the diminishing difference between consecutive loop current peaks as input pulses accumulate and thus the distinguishability of adjacent peaks. In the theory calculation of bias-current programming section of methods, this effect is neglected during theoretical analysis for computational simplicity. Due to the exponential decay effect in the LIF, the loop current peaks follow the recursive relation in reality,

$$I_k=\alpha I_{k-1}+\Delta I,$$

where $I_k$ is the loop current after the $k$-th input pulse, $\alpha$ is the decay factor $\alpha=e^{-\frac{1}{\tau f}}(<1)$, and $\Delta I$ is the instantaneous current increment generated by a pulse. The closed-form solution is as follows,

$$I_k=\Delta I\frac{1-\alpha^k}{1-\alpha}.$$

This finding suggests that loop current peak growth gradually slows down with the number of input pulses. Simulation of ten consecutive pulses shows that the difference between consecutive peaks decreases to only a few μA (Fig. 3b). Instrumental and thermal noise all contribute to the fluctuations in the observed peak currents, leading to variations in synaptic weight.

**Evaluation of Spatial Uniformity of the Neurons**

We employed the two-sample Kolmogorov-Smirnov test (K-S) test to evaluate the parameter distributions across 16 synapses within the network. Specifically, each synapse underwent independent parameter scanning yielding 2,352 measurements per synapse, totaling 37,632 measurements. Given this large sample size, traditional hypothesis tests based solely on p-values can become overly sensitive, flagging physically negligible deviations as statistically significant. In contrast, the K-S test focuses on the maximum distance between the empirical cumulative distribution functions (ECDFs), providing a robust metric for distributional consistency that is less sensitive to the effects of large-scale sampling. Within the K-S test, we used the synapse at coordinates (1,1) as the reference point to assess the consistency of the remaining 15 synapses. For each pair, we recorded the K-S statistic D, defined as the maximum vertical difference between their ECDFs. To derive a simple and sample-size-robust similarity metric, we defined

$$score = 1 - D,$$

where a value closer to 1 indicates stronger distributional agreement. As shown in Fig. 4d, the scores range from 0.772 to 0.916 (mean ≈ 0.85), indicating good spatial consistency in parameter distribution across the 15 synapses relative to the reference.

**SPINIC performance calculations.**

The energy consumption for a single synaptic operation ($E_{syn}$) is derived from the power model ($P_{syn} = 7.24\ \mu W$) and the ultrafast synaptic event duration ($T_{syn} \approx 444\ ps$ at 45 GHz). This yields an energy per synaptic operation (SOP) of:

$$E_{syn} = P_{syn} T_{syn} \approx 3.21\ \text{fJ/SOP}.$$

The computational throughput of the SPINIC neuromorphic core is quantified using the synaptic operations per second (SOPS) metric, where one synaptic operation corresponds to a single spike being processed by one synapse (i.e., weighted transmission from a pre-synaptic to a post-synaptic neuron). For an $N \times N$ synaptic array that contains $N^2$ programmable synapses, the peak SOPS is determined by the maximum operating frequency of the synaptic circuit, which is limited by the minimum time required to complete one synaptic operation, denoted $T_{syn}$. This parameter was extracted from transient simulations of the full synapse circuit under the worst-case loading conditions. Assuming all $N^2$ synapses operate in parallel at their maximum rate (a valid assumption in asynchronous SFQ logic), the peak throughput is given by

$$SOPS = \frac{N^2}{T_{syn}}.$$

Thus, for a 32×32 SPINIC core, its SOPS is:

$$SOPS_{32\times32} = \frac{32^2}{444 \times 10^{-12}\text{s}} \approx 2306.31 \times 10^9$$

For energy efficiency calculations, we extrapolate from the power consumption of a 4×4 network to a larger 32×32 configuration. The 4×4 network consumes 386.76 μW, equating to an average power consumption per processing element (PE) of

$$P_{PE} = \frac{386.76\ \mu\text{W}}{4^2} \approx 24.17\ \mu\text{W}.$$

This allows us to calculate the power consumption of a 32×32 network to be approximately as

$$P_{32\times32} = 24.17\ \mu\text{W} \times 32^2 \approx 24.75\ mW.$$

The ideal energy efficiency (EE) is obtained by the ratio of throughput to the power consumption as

$$\text{EE}_{ideal} = \frac{SOPS_{32\times32}}{P_{32\times32}} \approx 93{,}184\ \text{GSOPS/W}.$$

Considering the impact of refrigeration power consumption needed for practical superconducting circuits, the estimated energy efficiency can be derived as follows,

Considering the impact of the actual refrigeration power consumption required for superconducting circuits (which is approximately 300 times the circuit power consumption[3]), its estimated energy efficiency can be derived as follows:

$$\text{EE}_{real} = \frac{\text{EE}_{ideal}}{300} \approx 311\ GSOPS/W$$

Notice that the ratio of static to dynamic power consumption in our current RSFQ circuit is as follows,

$$\frac{P_S}{P_D} = \frac{I_B \times V_B}{f \times \Phi_0 \times I_B} = \frac{V_B}{f \times \Phi_0} = \frac{2.6mV}{45GHz \times 2.07 \times 10^{-15} Wb} \approx 27.91,$$

where $V_B$ is the circuit bias voltage, $f$ is the circuit operating frequency, and $\Phi_0$ is the value of a single magnetic flux quantum. By utilizing more energy-efficient ERSFQ technology[4] for building the same SPINIC, the primary source of power dissipation, i.e., the static power consumption in circuits, can be eliminated, leaving only dynamic power consumption. Under this scenario, the ideal and actual energy efficiency ratios of SPINIC can be inferred to improve to the following levels:

$$EE_{ideal}^{ERSFQ} = \mathrm{EE}_{ideal} \times (27.91 + 1) \approx 2{,}693{,}949\ \mathrm{GSOPS/W}$$

$$EE_{real}^{ERSFQ} = \mathrm{EE}_{real} \times (27.91 + 1) \approx 8962\ \mathrm{GSOPS/W}$$

**Potential Applications in Quantum-Classical Intelligence and Spiking Language Models**

One of the most pressing challenges in scaling superconducting quantum computers is real-time quantum error correction (QEC). Surface code decoding requires nanosecond-scale decision-making based on syndrome measurements, which is an ideal match with SPINIC's GHz-speed operation. Unlike FPGA-based decoders that introduce significant latency and power overhead, SPINIC can be monolithically or heterogeneously integrated with Nb-based qubit chips and operates in the same dilution refrigerator. Its analog programmability further allows adaptive tuning of decoding logic in response to time-varying noise environments, enabling intelligent, self-optimizing quantum error correction (QEC) schemes. Additionally, SPINIC could serve as a low-latency feedback controller for dynamic qubit calibration, gate optimization, or entanglement scheduling via reinforcement learning policies executed directly at cryogenic temperatures.

Beyond quantum applications, SPINIC opens new possibilities for ultra-efficient natural language processing through spiking large language models (Spiking LLMs). While current LLMs are computationally prohibitive due to massive parameter counts and dense attention mechanisms, recent advances in SNN-based transformers show promise in reducing energy consumption through sparse, event-driven computation. SPINIC's high-speed, low-energy neuron dynamics make it an ideal hardware substrate for deploying such models in latency-critical scenarios. For example, a SPINIC-based Spiking LLM could perform real-time speech-to-text translation, command recognition, or context-aware reasoning in edge devices with strict power budgets. Moreover, the temporal coding capabilities of SNNs that are natively supported by SPINIC can capture sequential dependencies more efficiently than frame-based ANNs, potentially improving inference fidelity while reducing computational load. Future work by incorporating integrating attention mechanisms and positional encoding into superconducting SNNs could unlock a new class of terascale spiking language processors.

**Current Limitations and Future Outlook**

While SPINIC demonstrates compelling performance in speed, energy efficiency, and accuracy, its deployment involves several practical considerations common to cryogenic computing platforms. To address the interconnect scaling challenge posed by individual bias lines in large-scale networks, we propose a hierarchical control architecture for future SPINIC development. We envision a scenario by implementing a matrix addressing scheme (similar to Word-line/Bit-line structures in semiconductor memories) combined with on-chip superconducting D/A converters or sample-and-hold circuits. This would allow for serial loading and parallel retention of parameters across the array. Furthermore, by adopting Cryo-CMOS co-design, multiplexing and demultiplexing logic can be integrated directly at the 4 K stage. This heterogeneous approach would dramatically reduce the number of physical I/O lines required between room temperature and the cryogenic stage, effectively mitigating the wiring bottleneck.

Steady advancement in cryocooler technology is expected to help reduce the operational footprint of superconducting neuromorphic systems and thus broaden their applicability. The interface between room-temperature control electronics and cryogenic circuits remains an engineering challenge, particularly for large-scale data transfer. However, this issue is actively being addressed through cryo-CMOS co-design and signal multiplexing techniques, benefiting a broader superconducting ecosystem. Integrating SPINIC with emerging cryogenic interface solutions could effectively mitigate I/O bottlenecks in future deployments.

Although current fabrication processes offer high yield for moderate-scale circuits, scaling to tens of thousands of neurons will benefit from continued improvements in multi-layer wiring and design automation tools. The modular, tile-based architecture of SPINIC, is inherently scalable, and ongoing progress in 3D integration and heterogeneous packaging provides a clear pathway toward larger networks. Current engineering challenges along this path are aligned with those lying in quantum and high-performance computing, suggesting that coordinated advancement in fabrication, packaging, and system integration will make SPINIC or similar architectures even more practical so as to play a growing role in next-generation intelligent systems.

Beyond efficient inference, SPINIC's intrinsic dual-timescale plasticity offers a foundation for on-chip in-situ learning. Future work will focus on integrating SFQ-based Spike-Timing-Dependent Plasticity (STDP) circuits that detect pre- and post-synaptic spike timing differences and convert this temporal information into feedback signals to dynamically modulate synaptic weights (encoded as bias currents). By establishing this closed-loop mechanism, SPINIC will evolve beyond offline PC-based training to achieve autonomous, real-time parameter adaptation directly on the hardware—realizing the vision of ultra-low-power, self-learning neuromorphic systems.